\pdfoutput=1
\documentclass[journal,10pt,twocolumn]{IEEEtran}

\usepackage{cite}
\ifCLASSINFOpdf
  \usepackage[pdftex]{graphicx}
\else
  \usepackage[dvips]{graphicx}
\fi
\usepackage{tabularx}
\usepackage{subfigure}
\usepackage{color}
\usepackage{algorithm}
\usepackage{algorithmicx}
\usepackage{algpseudocode}
\usepackage{amsfonts}
\usepackage{amsmath}
\usepackage{amssymb}
\usepackage{array}

\begin{document}

\title{Scattering Environment Aware Joint Multi-BS Channel Estimation and Localization with Clock Asynchronism}
\DeclareRobustCommand*{\IEEEauthorrefmark}[1]{%
    \raisebox{0pt}[0pt][0pt]{\textsuperscript{\footnotesize\ensuremath{#1}}}}
\author{Yani Chi.
\thanks{
Y.~Chi is with the  State Key Laboratory of Advanced Optical Communication Systems and Networks,  School of Electronics, Peking University. (e-mail: chiyani@caict.ac.cn).
}
}

\maketitle

% As a general rule, do not put math, special symbols, or citations
\begin{abstract}
Clock asynchronism between base stations (BSs) and users significantly degrades scatterer localization accuracy. To address this issue, this paper proposes a multi-BS joint channel estimation and localization scheme that leverages shared scatterer information among multiple BSs. First, channel modeling in the location domain is performed by exploiting the joint sparsity of multi-BS channels. Subsequently, a multi-BS scatterer association algorithm is designed based solely on angle-of-arrival (AoA) estimates. By leveraging the shared scatterers, the geometric relationships among the scatterers, BSs, and the UE enable the coarse estimation of the user location and timing offsets. Building upon these coarse estimates of scatterer positions, UE location, and timing offsets, an expectation-maximization (EM)-based framework is then employed. Specifically, the UE location and timing offsets are iteratively refined, while jointly enabling high-precision estimation of both scatterer locations and channel coefficients. Simulation results demonstrate that the proposed scheme achieves significant performance gains in both channel estimation and localization accuracy compared with baseline methods.
%The proposed method is structured into three distinct phases: Phase One: Individual MUSIC rough estimations are conducted at each base station. The input to this phase is the uplink signal received at each base station, with the output being the estimated angles of the surrounding scatterers for each base station. Phase Two: A novel scatterers associate algorithm is employed for multi-station target matching and rough localization. This phase takes as input the angles of the scatterers surrounding each base station and outputs the target matching results along with the localization of the scatterers.Phase Three: A refined estimation, known as Turbo-CS, is collaboratively performed across all base stations. The input for this phase is again the uplink signal received at each base station, and the output includes the channel map and the refined channel estimation. Our methodology leverages the spatial awareness of the scattering environment to achieve a joint channel estimation and localization that surpasses the limitations of single base station approaches. 
\end{abstract}

\begin{IEEEkeywords}
Channel estimation and localization, multi-BS, time offset.
\end{IEEEkeywords}
% \fi
\IEEEpeerreviewmaketitle

\section{Introduction}
With the rapid development of sixth-generation (6G) intelligent machine-type applications, such as intelligent vehicular networks, smart factories, and smart cities, both communication and sensing are indispensable for the flexible operation of autonomous machines~\cite{ref1}. The widespread deployment of massive multiple-input multiple-output (MIMO) and orthogonal frequency division multiplexing (OFDM) technologies ensures that communication signals in future wireless systems will exhibit superior resolution in both the delay and angle domains~\cite{ref2}, thereby enabling high-precision sensing using communication signals.

However, the clock asynchronism between the base station (BS) and user equipment (UE) introduces time offsets, which significantly degrade the accuracy of localization estimation in MIMO-OFDM systems~\cite{ref4}. Existing works typically rely on restrictive assumptions to address this issue. For instance,~\cite{ref5} investigated the single-BS localization problem under clock offset, assuming the presence of a line-of-sight (LoS) path along with at least one additional multipath component. Similarly,~\cite{ref6} considered a comparable setup that depends on multiple snapshots and the availability of a LoS path. However, these strong assumptions may not hold in practical scenarios, thereby limiting the applicability of such approaches.

In addition, some studies have explored using multi-BS systems to observe the surrounding scatterer environment from different perspectives to improve localization accuracy. In~\cite{ref7}, multi-BS systems cooperatively process multi-view data and jointly sense the entire environment, accounting for occlusion effects. In~\cite{ref8}, a location-based matching method is proposed to fuse sensing information from multiple BSs. 
In~\cite{ref9}, a physics-aware generative multi-view sensing framework was proposed, where the multi-BS channel is fused via a dedicated encoder and further leveraged by a diffusion model to achieve high-precision target reconstruction. However, existing multi-BS solutions typically rely on either prior knowledge of the UE location or the assumption of perfect clock synchronization, which enables direct alignment of observations across multiple BSs. 

It is worth noting that, in single-BS scenarios, when the UE location is unknown and timing offsets are present, the positions of scatterers become inherently ambiguous, rendering the localization problem ill-posed. In contrast, by leveraging multiple BSs, this ambiguity can be resolved under certain conditions, as will be detailed in Section II. 

In this letter, we propose a joint multi-BS channel estimation and localization scheme that mitigates time offsets by aligning multiple BSs through shared scatterers. First, we model the multi-BS system in the location domain by exploiting the partially common sparsity of the channels, enabling joint channel estimation and localization. Since absolute delays are unavailable, a multi-BS scatterer association method based solely on angle-of-arrival (AoA) information is developed. Based on the associated shared scatterers, coarse estimates of the UE location and timing offsets are obtained via the underlying geometric relationships. Building upon these coarse estimates, an expectation-maximization (EM)-based framework is further proposed, where the UE location and timing offsets are iteratively refined, while enabling high-precision estimation of both scatterer locations and channel coefficients by fully exploiting the structured sparsity of the multi-BS system.
\begin{figure}
	\centering
	\includegraphics[width=0.5\textwidth]{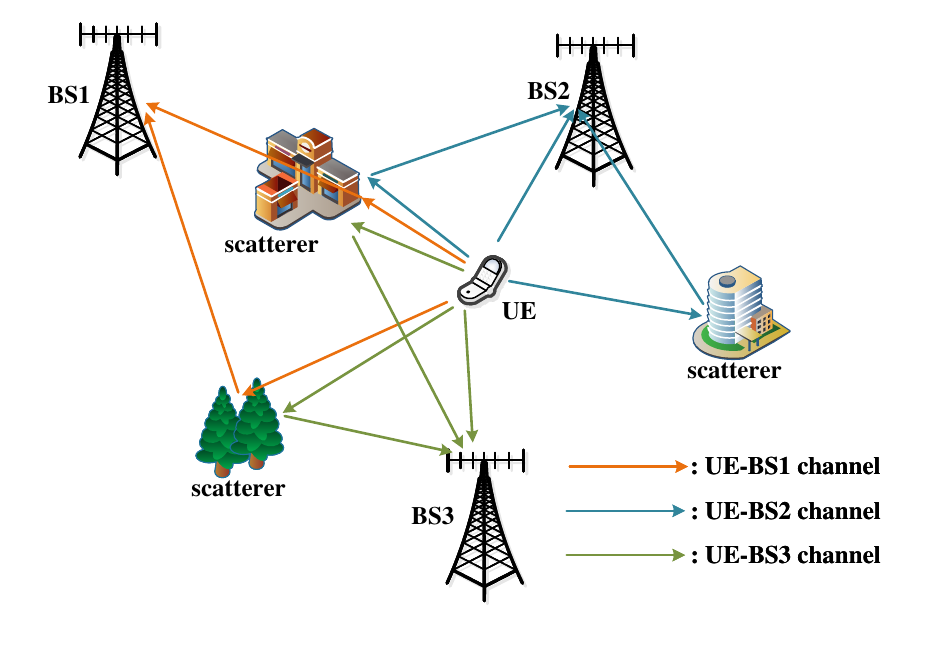}
	\caption{Multi-BS uplink channel model.}
	\label{Fig:model}
\end{figure}
 \section{System Model }\label{sec:system_model}
    \subsection{Channel Model}

We consider a TDD MIMO-OFDM system consisting of $K$ base stations (BSs), each equipped with $N = N_x N_y$ antennas, and a single-antenna user, as illustrated in Fig.~\ref{Fig:model}. Let $\mathcal{K}=\{1,\ldots,K\}$ denote the BS index set. The user is located at $\boldsymbol{p}_{\mathrm{UE}} = [p_{\mathrm{UE}}^x, p_{\mathrm{UE}}^y, p_{\mathrm{UE}}^z]^T$. The position of the $k$-th BS is known and denoted by $\boldsymbol{p}_{\mathrm{BS},k} = [p_{\mathrm{BS},k}^x, p_{\mathrm{BS},k}^y, p_{\mathrm{BS},k}^z]^T$, $k \in \mathcal{K}$. Assume that $L$ scatterers are distributed within a three-dimensional region $\mathcal{R}$. Let $\mathcal{L} = \{1, \ldots, L\}$ denote the index set of the scatterers, and $\boldsymbol{p}_{\mathrm{SC},l} = [p_{\mathrm{SC},l}^x, p_{\mathrm{SC},l}^y, p_{\mathrm{SC},l}^z]^T$ denote the position of the $l$-th scatterer. The subset of scatterers contributing to the propagation paths between the user and the $k$-th BS is denoted by $\mathcal{L}_k \subseteq \mathcal{L}$. Our objective is to estimate the uplink channels while sensing the scattering environment. On the $p$-th subcarrier $(1 \le p \le P)$, the user transmits a pilot symbol $u_p$. The received signal at the $k$-th BS is given by
\begin{equation}
\boldsymbol{y}_{k,p}=\boldsymbol{h}_{k,p}u_p+\boldsymbol{z}_{k,p},
\end{equation}
where $\boldsymbol{h}_{k,p}\in\mathbb{C}^N$ denotes the channel vector and $\boldsymbol{z}_{k,p}\sim\mathcal{CN}(\mathbf{0},(\sigma_{k,p})^2\boldsymbol{I})\in\mathbb{C}^N$ is additive white Gaussian noise.
Let $\theta_{k,l}$, $\phi_{k,l}$, and $\tau_{k,l}$ denote the azimuth angle, elevation angle, and propagation delay associated with the $l$-th scatterer, respectively. These parameters are determined by the relative geometry between the $k$-th BS and the $l$-th scatterer as
\begin{equation}
\theta_k(\boldsymbol{p}_{\mathrm{SC},l})=
\operatorname{atan2}\!\left(
p_{\mathrm{SC},l}^y-p_{\mathrm{BS},k}^y,\,
p_{\mathrm{SC},l}^x-p_{\mathrm{BS},k}^x
\right),
\end{equation}

\begin{equation}
\phi_k(\boldsymbol{p}_{\mathrm{SC},l}) =
\operatorname{atan2}\!\left(
p_{\mathrm{SC},l}^z - p_{\mathrm{BS},k}^z,\,
d_{k,l}^{xy}
\right),
\end{equation}

\begin{equation}
\tau_k(\boldsymbol{p}_{\mathrm{SC},l})=
\frac{\|\boldsymbol{p}_{\mathrm{BS},k}-\boldsymbol{p}_{\mathrm{SC},l}\|}{c}
+
\frac{\|\boldsymbol{p}_{\mathrm{UE}}-\boldsymbol{p}_{\mathrm{SC},l}\|}{c},
\end{equation}
where $d_{k,l}^{xy} = \sqrt{(p_{\mathrm{SC},l}^x - p_{\mathrm{BS},k}^x)^2 + (p_{\mathrm{SC},l}^y - p_{\mathrm{BS},k}^y)^2}$, $\|\cdot\|$ denotes the Euclidean norm and $c$ is the speed of light. For convenience, we stack the received signals across all subcarriers at the $k$-th BS as
\begin{equation}
\boldsymbol{y}_k=
\left[
(\boldsymbol{y}_{k,1})^T,\ldots,(\boldsymbol{y}_{k,P})^T
\right]^T
\in\mathbb{C}^{NP\times1}.
\end{equation}

The channel vector on subcarrier $p$ can be modeled as
\begin{equation}
\boldsymbol{h}_{k,p}=
\sum_{l\in\mathcal{L}_k}
x_{k,l}
e^{-j2\pi p f_0 \Delta\tau_k(\boldsymbol{p}_{\mathrm{SC},l})}
\boldsymbol{a}_k(\boldsymbol{p}_{\mathrm{SC},l}),
\end{equation}
where $x_{k,l}$ is the complex channel gain of the path from the user to the $k$-th BS via the $l$-th scatterer, and $\Delta\tau_k(\boldsymbol{p}_{\mathrm{SC},l})=\tau_k(\boldsymbol{p}_{\mathrm{SC},l})+\delta_k$ denotes the relative delay including the timing offset $\delta_k$. The timing offset is assumed to follow the prior distribution $\delta_k\sim\mathcal{N}(0,\sigma_\tau^2)$~\cite{ref3}. Moreover, $f_0$ denotes the subcarrier spacing, and $\boldsymbol{a}_k(\boldsymbol{p}_{\mathrm{SC},l})$ is the array steering vector of the $k$-th BS.
\subsection{Multi-BS Joint Sparsity Model}
Due to the physical propagation characteristics of electromagnetic waves, each BS can only receive signals reflected from a subset of environmental objects. Therefore, each BS has its own sensing region. 
This property of multi-BS channels can be modeled as partially common sparsity in the location domain. We define a three-dimensional (3D) dynamic location-domain grid 
\[
\boldsymbol{q}=\{\boldsymbol{q}_1,\ldots,\boldsymbol{q}_Q\}\subset\mathcal{R},
\]
consisting of $Q$ candidate Cartesian locations. Let $\mathcal{Q}=\{1,\ldots,Q\}$ denote the grid index set. The $q$-th grid point $\boldsymbol{q}_q$ can also be represented from the viewpoint of the $k$-th BS by the BS-dependent angular-range parameters $(\theta_{k,q},\phi_{k,q},r_{k,q})$.

For the $k$-th BS, we define the following two basis matrices
$\boldsymbol{A}_k\in\mathbb{C}^{N\times Q}$ and 
$\boldsymbol{b}_{k,p}\in\mathbb{C}^{Q\times1}$ as
\begin{equation}\label{7}
\boldsymbol{A}_k=
\left[
\boldsymbol{a}_k(\boldsymbol{q}_1),
\cdots,
\boldsymbol{a}_k(\boldsymbol{q}_Q)
\right],
\end{equation}

\begin{equation}\label{8}
\boldsymbol{b}_{k,p}(\boldsymbol{p}_{\mathrm{UE}},\delta_k)=
\left[
e^{-j2\pi p f_0 \Delta\tau_k(\boldsymbol{q}_1)},
\cdots,
e^{-j2\pi p f_0 \Delta\tau_k(\boldsymbol{q}_Q)}
\right]^T .
\end{equation}
Here, $\Delta\tau_k(\boldsymbol{q}_q)$ is shorthand for $\tau_k(\boldsymbol{q}_q)+\delta_k$, where $\tau_k(\boldsymbol{q}_q)$ is obtained from the geometric delay expression by replacing $\boldsymbol{p}_{\mathrm{SC},l}$ with the candidate location $\boldsymbol{q}_q$.

Then, the sparse representation of the NLoS channel vector at the $p$-th subcarrier of the $k$-th BS is given by
\begin{equation}
\boldsymbol{h}_{k,p}^{\mathrm{NLoS}}=
\boldsymbol{A}_k
\operatorname{Diag}\!\left(
\boldsymbol{b}_{k,p}(\boldsymbol{p}_{\mathrm{UE}},\delta_k)
\right)
\boldsymbol{x}_k,
\end{equation}
where $\boldsymbol{x}_k$ is a sparse vector with only $L_k$ nonzero elements corresponding to $L_k$ scatterers. 
The $q$-th element of $\boldsymbol{x}_k$, denoted by $x_{k,q}$, represents the complex channel gain of the scatterer located at the $q$-th grid point, which follows an i.i.d. complex Gaussian distribution with zero mean and unit variance.

Then, we introduce a three-layer sparse prior model to capture the joint sparse structure of the multi-BS channels. Let 
\[
\boldsymbol{s}^k=[s_1^k,\ldots,s_Q^k]^T, \quad q\in\mathcal{Q},
\]
denote the channel support vector associated with the $k$-th BS, where $s_q^k\in\{-1,1\}$. 
In particular, $s_q^k=1$ indicates that there exists a communication scatterer around the $q$-th AoA grid $\theta_q$. 
Accordingly, the support set
\[
\Omega_k \triangleq \{q: s_q^k=1\}
\]
represents the set of coarse AoAs associated with the user at the $k$-th BS.

Conditioned on the support variables $s_q^k$, the channel coefficients $x_q^k$ are assumed to be independent of the conditional prior
\begin{equation}
p(x_q^k|s_q^k)
=
\delta(s_q^k+1)\delta(x_q^k)
+
\delta(s_q^k-1)
\mathcal{CN}(x_q^k;0,\sigma_q^2),
\quad q\in\mathcal{Q}.
\end{equation}

To model the partially common sparsity across multiple BSs, we introduce a global support vector
\[
\boldsymbol{s}=[s_1,\ldots,s_Q]^T.
\]
The joint distribution of the support variables is given by
\begin{equation}
p(\boldsymbol{s},\boldsymbol{s}^1,\ldots,\boldsymbol{s}^K)
=
p(\boldsymbol{s})
\prod_{k\in\mathcal{K}}
p(\boldsymbol{s}^k|\boldsymbol{s})
=
p(\boldsymbol{s})
\prod_{k\in\mathcal{K}}
\prod_{q\in\mathcal{Q}}
p(s_q^k|s_q).
\end{equation}
The conditional distribution is defined as
\begin{equation}
p(s_q^k|s_q)
=
\delta(s_q+1)\delta(s_q^k+1)
+
\delta(s_q-1)
\Big(
\rho_k\,\delta(s_q^k-1)
+
(1-\rho_k)\delta(s_q^k+1)
\Big),
\end{equation}
where $\rho_k$ denotes the probability that $s_q^k=1$ conditioned on $s_q=1$, characterizing the overlap between shared scatterers and BS-specific scatterers. The global support vector $\boldsymbol{s}$ follows an i.i.d. prior
\begin{equation}
p(\boldsymbol{s})
=
\prod_{q\in\mathcal{Q}} p(s_q),
\end{equation}
where the initialization probabilities are set as $p(s_q=1)=\lambda$ and $p(s_q=-1)=1-\lambda$, with $\lambda$ controlling the sparsity level.

Finally, the joint prior distribution of the support variables and channel coefficients is given by
\begin{align}
&p(\boldsymbol{s},\boldsymbol{s}^1,\ldots,\boldsymbol{s}^K,
\boldsymbol{x}^1,\ldots,\boldsymbol{x}^K)
\\
&=
p(\boldsymbol{s})
\prod_{k\in\mathcal{K}}
\prod_{q\in\mathcal{Q}}
p(s_q^k|s_q)
\prod_{k\in\mathcal{K}}
\prod_{q\in\mathcal{Q}}
p(x_q^k|s_q^k).
\end{align}

\subsection{Multi-BS Joint Channel Estimation and Localization Model}

The received signal at the $k$-th BS can be expressed as
\begin{equation}
\boldsymbol{y}_k=
\boldsymbol{\Phi}_k(\boldsymbol{p}_{\mathrm{UE}},\delta_k)\boldsymbol{x}_k
+\boldsymbol{z}_k .
\end{equation}

where
\[
\boldsymbol{y}_k=
[(\boldsymbol{y}_{k,1})^T,\ldots,(\boldsymbol{y}_{k,P})^T]^T
\in\mathbb{C}^{NP\times1},
\]
\[
\boldsymbol{x}_k=
[x_{k,1},\ldots,x_{k,Q}]^T
\in\mathbb{C}^{Q\times1},
\]
\[
\boldsymbol{z}_k=
[(\boldsymbol{z}_{k,1})^T,\ldots,(\boldsymbol{z}_{k,P})^T]^T
\in\mathbb{C}^{NP\times1}.
\]

The sensing matrix is given by
\begin{equation}
\boldsymbol{\Phi}_k(\boldsymbol{p}_{\mathrm{UE}},\delta_k)
=
\begin{bmatrix}
u_1\boldsymbol{A}_k\operatorname{diag}\!\big(\boldsymbol{b}_{k,1}(\boldsymbol{p}_{\mathrm{UE}},\delta_k)\big)\\
\vdots\\
u_P\boldsymbol{A}_k\operatorname{diag}\!\big(\boldsymbol{b}_{k,P}(\boldsymbol{p}_{\mathrm{UE}},\delta_k)\big)
\end{bmatrix}
\in\mathbb{C}^{NP\times Q}.
\end{equation}

The goal is to jointly estimate the channel coefficients 
$\boldsymbol{x}=\{\boldsymbol{x}_k\}_{k\in\mathcal{K}}$, 
the clock offsets 
$\boldsymbol{\delta}=[\delta_1,\ldots,\delta_K]$, 
and the user position $\boldsymbol{p}_{\mathrm{UE}}$ from the observations $\boldsymbol{y}$. Once $\boldsymbol{p}_{\mathrm{UE}}$ and $\boldsymbol{\delta}$ are obtained, the objective is to compute the conditional marginal posterior distributions
\[
p(x_q^k|\boldsymbol{y},\boldsymbol{p}_{\mathrm{UE}},\boldsymbol{\delta}),
\quad
p(s_q^k|\boldsymbol{y},\boldsymbol{p}_{\mathrm{UE}},\boldsymbol{\delta}),
\]
for all $q\in\mathcal{Q}$ and $k\in\mathcal{K}$. However, due to the presence of loop structures in the factor graph of the underlying probabilistic model, the exact computation of these marginal posterior distributions is intractable. To address this issue, we develop an EM-based framework to iteratively refine scatterer locations and channel coefficients by exploiting the partially common sparsity of the multi-BS system.

In the single-BS scenario, the parameters 
$\{\theta_1(\boldsymbol{p}_{\mathrm{SC},l})\}_{l=1}^L$, 
$\{\phi_1(\boldsymbol{p}_{\mathrm{SC},l})\}_{l=1}^L$, 
and 
$\{\Delta\tau_1(\boldsymbol{p}_{\mathrm{SC},l})\}_{l=1}^L$
can be estimated from the received signals. However, the scatterer positions 
$\{\boldsymbol{p}_{\mathrm{SC},l}\}_{l=1}^L$, 
the user position $\boldsymbol{p}_{\mathrm{UE}}$, 
and the clock offset $\delta_1$ must also be estimated, leading to an underdetermined system with $3L+4$ unknown variables but only $3L$ equations.

\begin{figure}[t]
\centering
\includegraphics[width=0.48\textwidth]{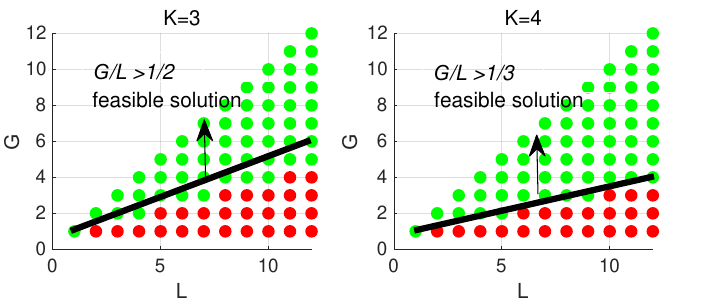}
\caption{Conditions for feasible solutions in the multi-BS scenario.}
\label{Fig:feasible}
\end{figure}

In contrast, in the multi-BS scenario, shared scatterers observed from different BSs provide additional constraints, resulting in an overdetermined system. Specifically, the system contains $3L+3+K$ unknowns. If $G$ denotes the number of shared scatterers, then one BS can be regarded as the reference view and each of the remaining $K-1$ BSs contributes additional cross-BS consistency constraints. A necessary degree-of-freedom condition for a feasible coarse solution is therefore $3G(K-1)\ge 3L+3+K$. As illustrated in Fig.~\ref{Fig:feasible}, when three BSs are available, the required shared-scatterer ratio approaches $G/L>1/2$, whereas for four BSs the requirement approaches $G/L>1/3$.

\subsection{Angle-Only Multi-BS Scatterer Association}

The scatterer positions cannot be determined directly from the estimated delays due to clock offsets. Consequently, the scatterers observed by different BSs cannot be matched by simply minimizing their spatial distances. Moreover, because the BSs are located at different positions, each BS may observe a different subset of environmental scatterers, further complicating association across BSs. To address this issue, we propose a scatterer association algorithm that relies solely on AoA information. We assume that coarse estimates of the scatterer AoAs, $\hat{\theta}_{k,l}$ and $\hat{\phi}_{k,l}$, as well as the relative delay $\Delta\hat{\tau}_{k,l}$, can be obtained using the 3-D MUSIC algorithm~\cite{ref9}. The main steps of the proposed algorithm are as follows.

\begin{figure}
	\centering
	\includegraphics[width=0.4\textwidth]{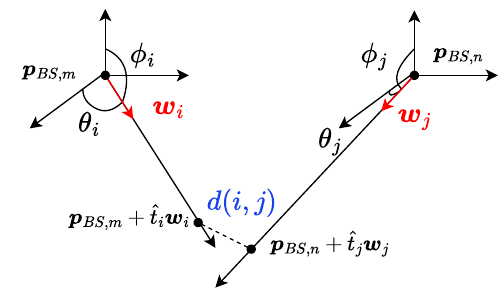}
 \caption{The distance between two rays.}
 \label{Fig:ray-distance}
\end{figure}
\textit{1) Definition of the distance between two AoA rays:}

Consider the $i$-th AoA observed at the $m$-th BS and the $j$-th AoA observed at the $n$-th BS, denoted by $(\hat{\theta}_i,\hat{\phi}_i)$ and $(\hat{\theta}_j,\hat{\phi}_j)$, respectively. 
These AoAs define two rays

\begin{equation}
l_i:\boldsymbol{p}_{BS,m}+t_i\boldsymbol{\hat w}_i,\quad
l_j:\boldsymbol{p}_{BS,n}+t_j\boldsymbol{\hat w}_j .
\end{equation}

The unit direction vectors are

\begin{equation}
\boldsymbol{\hat w}_i=
\begin{bmatrix}
\cos\hat{\theta}_i\cos\hat{\phi}_i \\
\sin\hat{\theta}_i\cos\hat{\phi}_i \\
\sin\hat{\phi}_i
\end{bmatrix},
\quad
\boldsymbol{\hat w}_j=
\begin{bmatrix}
\cos\hat{\theta}_j\cos\hat{\phi}_j \\
\sin\hat{\theta}_j\cos\hat{\phi}_j \\
\sin\hat{\phi}_j
\end{bmatrix}.
\end{equation}

The distance between the two rays is obtained by

\begin{equation}
\min_{t_i,t_j}
\left\|
\boldsymbol{d}_{mn}
+t_i\boldsymbol{\hat w}_i
-t_j\boldsymbol{\hat w}_j
\right\|^2 ,
\end{equation}
where $\boldsymbol{d}_{mn}=\boldsymbol{p}_{BS,m}-\boldsymbol{p}_{BS,n}.$ The LS solution can be written as

\begin{align}
\hat t_i &=
\frac{-\beta(\boldsymbol{d}_{mn}^T\boldsymbol{\hat w}_i)
+\gamma(\boldsymbol{d}_{mn}^T\boldsymbol{\hat w}_j)}
{\alpha\beta-\gamma^2},\\
\hat t_j &=
\frac{-\gamma(\boldsymbol{d}_{mn}^T\boldsymbol{\hat w}_i)
+\alpha(\boldsymbol{d}_{mn}^T\boldsymbol{\hat w}_j)}
{\alpha\beta-\gamma^2},
\end{align}
where $
\alpha=\boldsymbol{\hat w}_i^T\boldsymbol{\hat w}_i,\quad
\beta=\boldsymbol{\hat w}_j^T\boldsymbol{\hat w}_j,\quad
\gamma=\boldsymbol{\hat w}_i^T\boldsymbol{\hat w}_j .
$ Thus, the minimum distance between the two rays is

\begin{equation}
d(i,j)=
\begin{cases}
\|\boldsymbol{d}_{mn}
+\hat t_i\boldsymbol{\hat w}_i
-\hat t_j\boldsymbol{\hat w}_j\|,
& \hat t_i>0,\hat t_j>0,
\\
\|\boldsymbol{d}_{mn}
+\hat t_i\boldsymbol{\hat w}_i\|,
& \hat t_i>0,\hat t_j\le0,
\\
\|\boldsymbol{d}_{mn}
-\hat t_j\boldsymbol{\hat w}_j\|,
& \hat t_i\le0,\hat t_j>0,
\\
\|\boldsymbol{d}_{mn}\|,
& \hat t_i\le0,\hat t_j\le0 .
\end{cases}
\end{equation}

\begin{algorithm}[t]
\caption{Multi-BS Scatterer Association}
\label{alg:scatterer_association}
\begin{algorithmic}[1]

\State \textbf{Input:} scatterer index sets $\{\mathcal{L}_k\}_{k=1}^K$, distance threshold $d_{\mathrm{thr}}$
\State \textbf{Initialization:} 
$\mathcal{L}_{\mathrm{all}}=\bigcup_{k=1}^K \mathcal{L}_k$, 
$\mathcal{G}=\varnothing$, $g=1$

\While{$\mathcal{L}_{\mathrm{all}}\neq\varnothing$}

    \State Select an element $u\in\mathcal{L}_{\mathrm{all}}$
    \State $\mathcal{G}_g \leftarrow \{u\}$

    \For{each $v\in\mathcal{L}_{\mathrm{all}}\setminus\{u\}$}

        \State $flag \leftarrow \textbf{true}$

        \For{each $w\in\mathcal{G}_g$}

            \If{$d(v,w) < d_{\mathrm{thr}}$}
                \State $flag \leftarrow \textbf{false}$
                \State \textbf{break}
            \EndIf

        \EndFor

        \If{$flag$}
            \State $\mathcal{G}_g \leftarrow \mathcal{G}_g \cup \{v\}$
        \EndIf

    \EndFor

    \State $\mathcal{L}_{\mathrm{all}}
    \leftarrow
    \mathcal{L}_{\mathrm{all}}\setminus \mathcal{G}_g$

    \State $\mathcal{G} \leftarrow \mathcal{G}\cup\{\mathcal{G}_g\}$

    \State $g \leftarrow g+1$

\EndWhile

\State \textbf{Output:} cluster set $\mathcal{G}$ and number of shared scatterers $\hat G=g-1$

\end{algorithmic}
\end{algorithm}

\textit{2) Multi-BS Alignment Based on Shared Scatterers:}

We propose a multi-BS scatterer association algorithm that relies solely on AoA information. The core idea is to group scatterers observed by different BSs that correspond to the same physical object by exploiting the geometric consistency of their AoAs. Specifically, scatterers that are in close proximity (as defined in (20)) are merged into a single cluster, and the procedure is summarized in Algorithm~1. Based on the clustering results, shared scatterers across multiple BSs can be identified, which further helps eliminate false alarms.

Define $\boldsymbol{P}_{\mathrm{BS}}=[\boldsymbol{p}_{\mathrm{BS},1},\ldots,\boldsymbol{p}_{\mathrm{BS},K}]\in\mathbb{R}^{3\times K}$, $\boldsymbol{\theta}_g=[\theta_1(\boldsymbol{p}_{\mathrm{SC},l}),\ldots,\theta_K(\boldsymbol{p}_{\mathrm{SC},l})]^T\in\mathbb{R}^{K}$, and $\boldsymbol{\phi}_g=[\phi_1(\boldsymbol{p}_{\mathrm{SC},l}),\ldots,\phi_K(\boldsymbol{p}_{\mathrm{SC},l})]^T\in\mathbb{R}^{K}$, where $l\in\mathcal{G}_g$. The position of the $g$-th shared scatterer is estimated by solving
\begin{equation}
 \underset{\boldsymbol{p}_{\mathrm{SC},g}\in\mathbb{R}^3}{\min}
 \left\| \mathbf{G}_g \boldsymbol{p}_{\mathrm{SC},g} - \mathbf{h}_g \right\|^2,
\end{equation}
where $\mathbf{G}_g\in\mathbb{R}^{2K\times3}$ and $\mathbf{h}_g\in\mathbb{R}^{2K\times1}$ are given by
$$
\begin{aligned}
& 
\mathbf{G}_g=\begin{bmatrix}
-\sin(\boldsymbol{\theta}_g) & \cos(\boldsymbol{\theta}_g) & \mathbf{0} \\
-\sin(\boldsymbol{\phi}_g)\cos(\boldsymbol{\theta}_g) &
-\sin(\boldsymbol{\phi}_g)\sin(\boldsymbol{\theta}_g) &
\cos(\boldsymbol{\phi}_g)
\end{bmatrix}, \\
& 
\mathbf{h}_g=
\left(
\mathbf{1}_{1\times3}
\left(
\mathbf{G}_g^T \odot \boldsymbol{P}_{\mathrm{BS}}
\right)
\right)^T.
\end{aligned}
$$

It can be observed that the above formulation corresponds to a standard linear least-squares (LS) problem. Therefore, the estimate of the shared scatterer position can be obtained in closed form as
\begin{equation}
\boldsymbol{\hat{p}}_{SC,l}=\left(\mathbf{G}_g^T  \mathbf{G}_g\right)^{-1} \mathbf{G}_g^T \mathbf{h}_g, l \in \mathcal{G}_g.
\end{equation}

\subsection{Multi-BS Joint User Localization and Time Offset Estimation}

By leveraging observations from multi-BSs, the accuracy of user position estimation can be significantly improved. In particular, the proposed localization algorithm does not rely on the existence of a LoS path. Instead, cooperation among multiple BSs effectively resolves the ambiguities that are inherent in single-BS localization schemes. According to the equation (4), given $\boldsymbol{{p}}_{B\!S, k}$, $\boldsymbol{{p}}_{S\!C, l}$ and $\Delta \tau_{k,l}(\boldsymbol{{p}}_{S\!C, l}), l \in \mathcal{G}_g$, we have
\begin{equation}
\left\|\boldsymbol{{p}}_{U\!E}-\boldsymbol{{p}}_{S\!C, l}\right\|=\Delta r_{k,l}(\delta_k), l \in  \mathcal{G}_g, k \in \mathcal{K}
\end{equation}
where $\Delta r_{k,l}(\delta_k)=(\Delta \tau_{k}(\boldsymbol{{p}}_{S\!C, l})-\delta_k)c-\left\|\boldsymbol{{p}}_{B\!S, k}-\boldsymbol{{p}}_{S\!C, l}\right\|$ is the distance from the user to the $l$-th scatterer. The integration of all observations of the $g$-th shared scatterer can be represented as
\begin{equation}
\mathbf{1}_{GK \times 1}\boldsymbol{{p}}_{U\!E}^T \boldsymbol{{p}}_{U\!E} + \boldsymbol{u}-2  \boldsymbol{{p}} \boldsymbol{{p}}_{U\!E}=\boldsymbol{v},
\end{equation}
where $\boldsymbol{{p}}=[\boldsymbol{p}_1; \ldots; \boldsymbol{p}_{\hat{G}}] \in \mathbb{C}^{\hat{G}K \times 3}$ with $\boldsymbol{{p}}_g=\mathbf{1}_{K \times 1}\boldsymbol{\hat{p}}_{S\!C, g}^T \in \mathbb{C}^{K \times 3}$, $\boldsymbol{{u}}=[\boldsymbol{u}_1; \ldots; \boldsymbol{u}_{\hat{G}}] \in \mathbb{C}^{\hat{G}K \times 3}$ with $\boldsymbol{u}_g= \mathbf{1}_{K \times 1}\left\|\boldsymbol{{p}}_{S\!C, g}^T\right\|^2 \in \mathbb{C}^{K \times 3}$, and $\boldsymbol{{v}}=[\boldsymbol{v}_1; \ldots; \boldsymbol{v}_{\hat{G}}] \in \mathbb{C}^{\hat{G}K \times 3}$ with $\boldsymbol{v}_g=[\Delta r_{1,g}^2, \ldots \Delta r_{K,g}^2]^T \in \mathbb{C}^{K \times 1}$. 
However, due to the presence of errors in the estimated location of the scatterer, (24) can be transformed to minimize the following function:
\begin{equation}
 f(\boldsymbol{{p}}_{U\!E})=\|\mathbf{1}_{\hat{G}K \times 1}\boldsymbol{{p}}_{U\!E}^T \boldsymbol{{p}}_{U\!E} + \boldsymbol{\hat{u}}-2  \boldsymbol{\hat{p}} \boldsymbol{{p}}_{U\!E}-\boldsymbol{\hat{v}}\|^2,
\end{equation}
This problem is difficult to solve in closed form, so we update  $\boldsymbol{{p}}_{U\!E}$ using the gradient descent method, and $\boldsymbol{{p}}_{U\!E}$ at the $t$-th iteration is updated by

\begin{equation}
\boldsymbol{{p}}_{U\!E}^{(t+1)}=\boldsymbol{{p}}_{U\!E}^{(t)}-\epsilon \nabla f(\boldsymbol{{p}}_{U\!E}),
\end{equation}
where $\nabla f(\boldsymbol{{p}}_{U\!E}) = 2(\mathbf{1}_{\hat{G}K \times 1}\boldsymbol{{p}}_{U\!E}^T-\boldsymbol{\hat{p}})^T(\mathbf{1}_{\hat{G}K \times 1}\boldsymbol{{p}}_{U\!E}^T \boldsymbol{{p}}_{U\!E} + \boldsymbol{\hat{u}}-2  \boldsymbol{\hat{p}} \boldsymbol{{p}}_{U\!E}-\boldsymbol{\hat{v}})$ is the gradient of $f(\boldsymbol{{p}}_{U\!E})$, and $\epsilon$ is is a constant representing the learning rate. 

According to (24), the problem of the estimation of $\delta_k$ with given $\boldsymbol{{p}}_{U\!E}$ can be formulated as
\begin{equation}
 \underset{\delta_k}{ \min} \left\| \mathbf{1}_{\hat{G} \times 1}\delta_k-\boldsymbol{\hat{\gamma}}_k\right \|^2,
\end{equation}
where $\boldsymbol{\hat{\gamma}}_k=\Delta \tau_{k}(\boldsymbol{{p}}_{S\!C, l})-\frac{\|\boldsymbol{{p}}_{U\!E}-\boldsymbol{{p}}_{S\!C, l}\|}{c}-\frac{\|\boldsymbol{{p}}_{B\!S, k}-\boldsymbol{{p}}_{S\!C, l}\|}{c}$.  Applying the LS optimization to (29), the estimation of the time offset of the $k$-th BS can be expressed as

\begin{equation}
\hat{\delta}_k = (\mathbf{1}_{\hat{G} \times 1}^T\mathbf{1}_{\hat{G} \times 1})^{-1}\mathbf{1}_{\hat{G} \times 1}^T\boldsymbol{\hat{\gamma}}_k=\frac{1}{\hat{G}}\sum_{i=1}^{\hat{G}} \hat{\gamma}_{k,i}.
\end{equation}
where $\hat{\gamma}_{k,i}$ is the $k$-th element of $\boldsymbol{\hat{\gamma}}_k$. Thus, the algorithm for jointly estimating the user's location and time offset involves an iterative estimation of $\delta_k$ and $\boldsymbol{{p}}_{U\!E}$, as detailed in Algorithm 2. After the above calculation, we have the coarse estimation of scatterer AoA $\hat{\theta}_k\left(\boldsymbol{{p}}_{S\!C, l}\right)$, $\hat{\phi}_k\left(\boldsymbol{{p}}_{S\!C, l}\right)$, absolute delay $\hat{\tau}_{k}\left(\boldsymbol{{p}}_{S\!C, l}\right)$ and user location $\boldsymbol{\hat{p}}_{U\!E}$, where $\hat{\tau}_{k}\left(\boldsymbol{{p}}_{S\!C, l} \right)={\Delta} \hat{\tau}_k(\boldsymbol{{p}}_{S\!C, l})+{\hat{\delta}_k}$. According to the above information, we have the estimation of the location of all the scatterers $\boldsymbol{\hat{\xi}} = \bigcup_{k} \left\{ \boldsymbol{\hat{p}}_{S\!C, l} \,\middle|\, l \in \mathcal{L}_k \right\}$.

 \begin{figure}[t]
      \centering
      \includegraphics[width=3in]{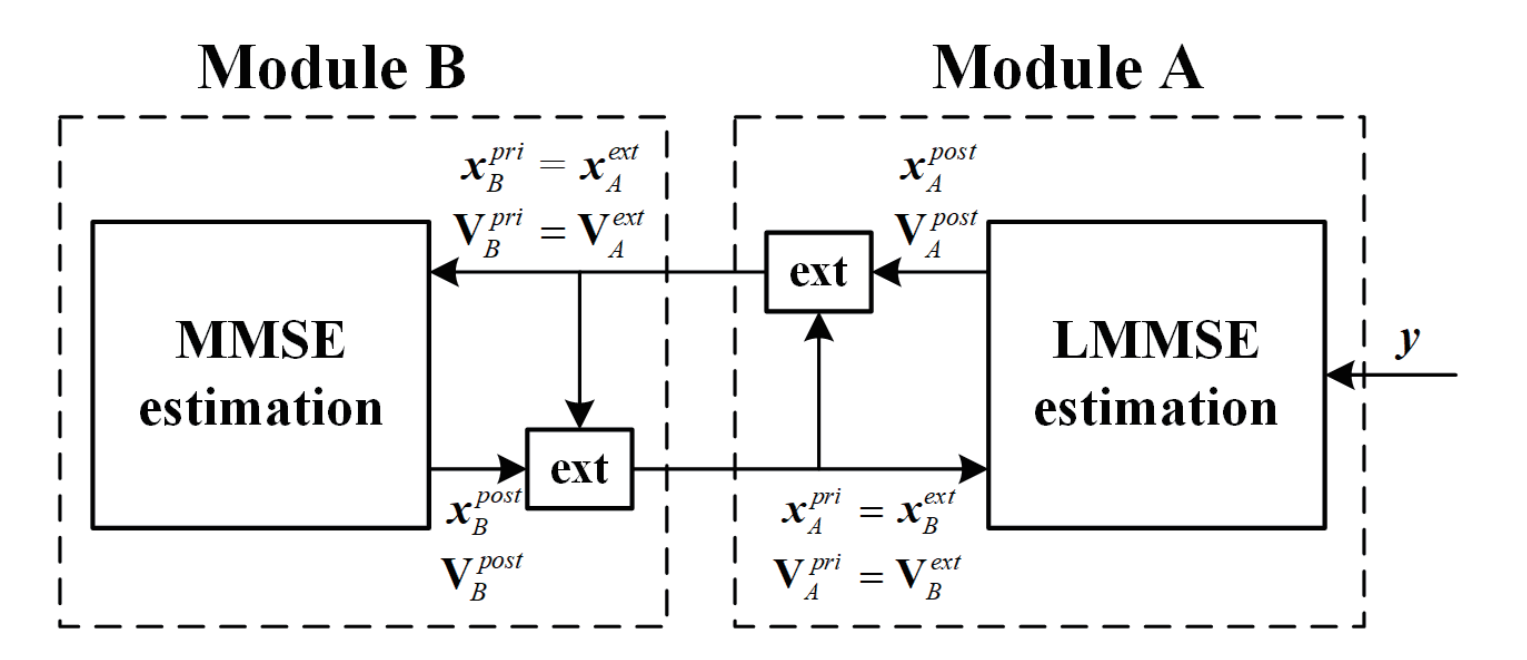}
      \caption{Illustration of the turbo approach.}
      \label{figure_1}
    \end{figure}
    \begin{figure}[t]
      \centering
      \includegraphics[width=3in]{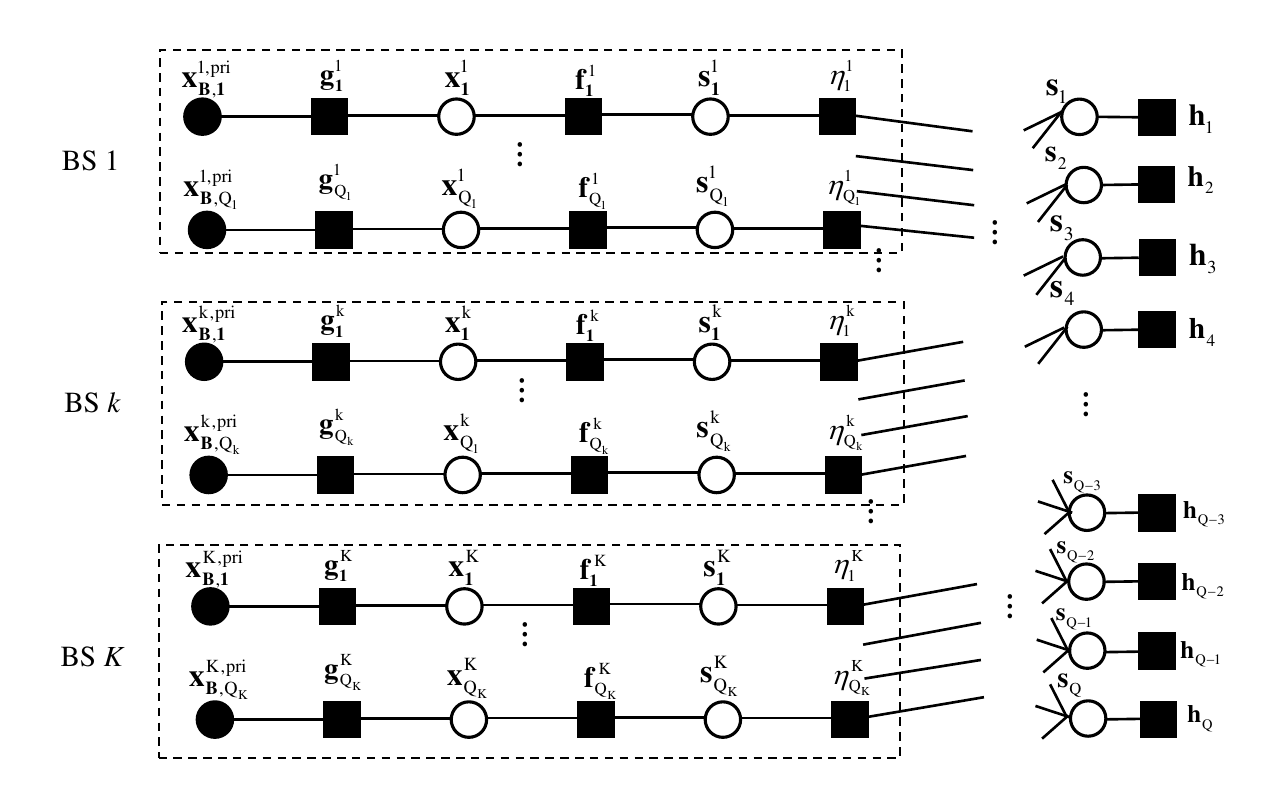}
      \caption{ The factor graph of the joint distribution of all variables.}
      \label{figure_factor_graph}
    \end{figure}

\begin{figure*}

	\setlength{\abovecaptionskip}{-5pt}
	\setlength{\belowcaptionskip}{-10pt}
	\centering
	\begin{minipage}[t]{0.33\linewidth}
		\centering
    \includegraphics[width=2.5in]{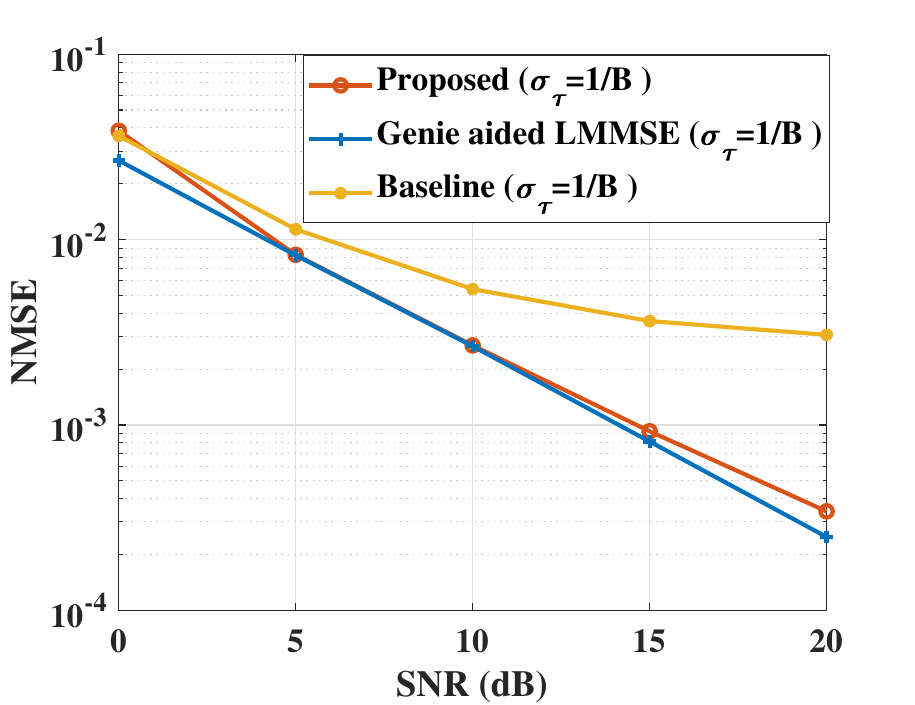}
		\caption{Channel estimation performance.}
		\label{figure1}
	\end{minipage}%
	\begin{minipage}[t]{0.33\linewidth}
		\centering

      \includegraphics[width=2.5in]{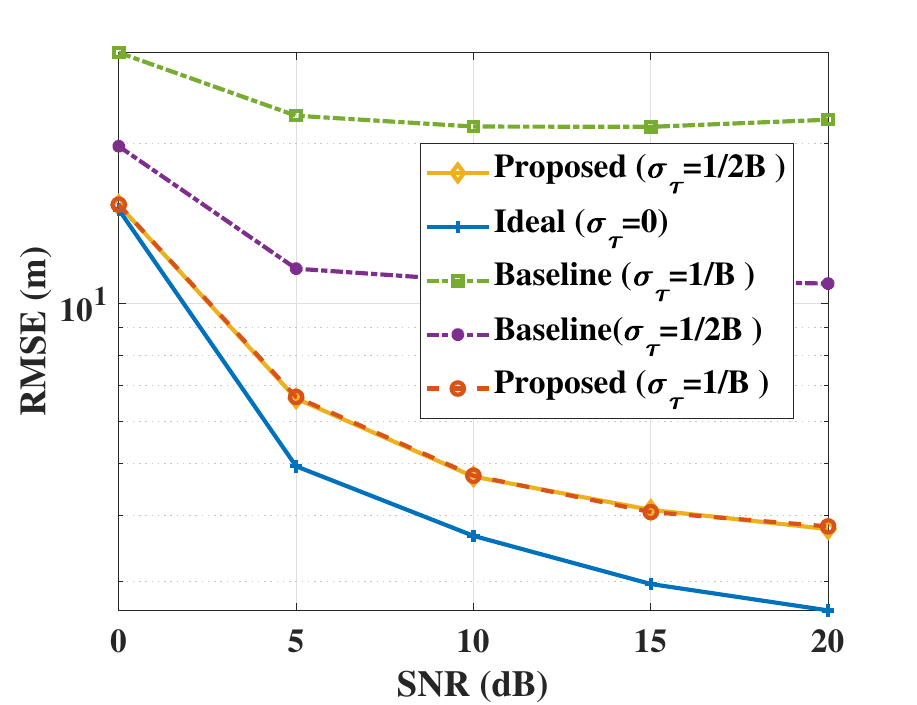}
		\caption{Scatterer localization performance.}

		\label{figure2}
	\end{minipage}
	\begin{minipage}[t]{0.33\linewidth}
		\centering
   \includegraphics[width=2.5in]{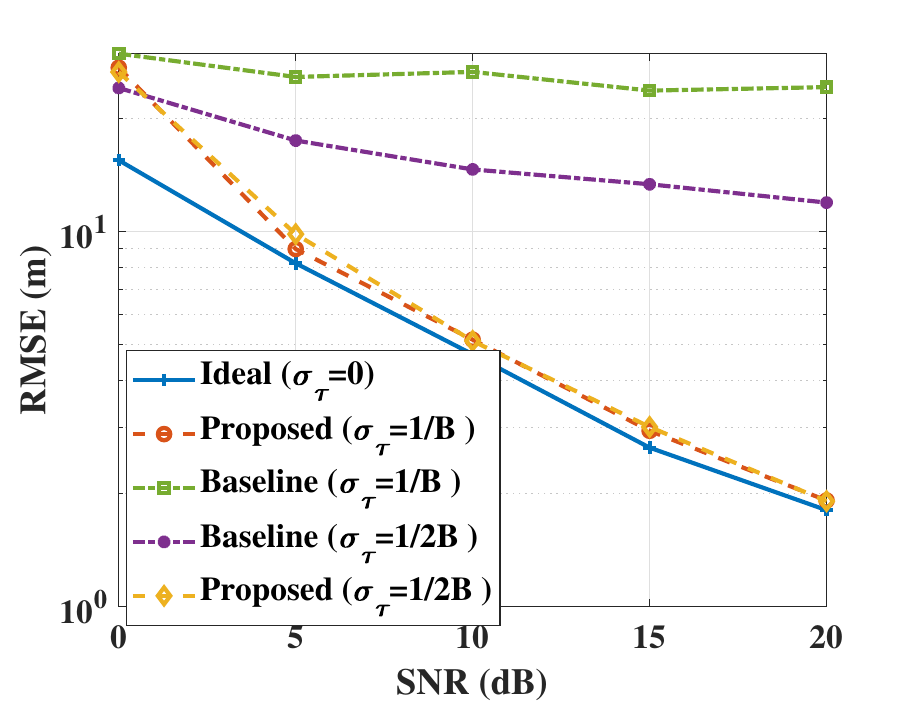}
		\caption{User localization performance.}
		\label{figure3}
	\end{minipage}
\end{figure*}

\begin{algorithm}
\caption{Joint User's Location and Time Offset Estimation}
\label{alg:distance_between_lines}
\begin{algorithmic}[1]
\State \textbf{Initialization}: Set initial $\boldsymbol{p}_{U\!E}^{(t)}$, $\delta_k^{(t)}$,  and iteration number $t=0$.
\State Update $\boldsymbol{p}_{U\!E}^{(t)}$ by (28) with given $\delta_k$
\State Update $\delta_k^{(t)}, \forall k \in \mathcal{K}$ by (30) with given $\boldsymbol{p}_{U\!E}$
\State Set $t\gets t+1$ and repeat the above until the convergence criteria are met.

\end{algorithmic}
\end{algorithm}

\section{Multi-BS Joint Channel Estimation and Localization}
\subsection{EM-Based Framework}

Based on the prior information of the system, denoted by
$\boldsymbol{\zeta}=\{\boldsymbol{\delta},\boldsymbol{p}_{\mathrm{UE}},\boldsymbol{\hat{\xi}}\}$,
we aim to perform joint channel estimation and localization. For given $\boldsymbol{y}$, $\boldsymbol{\delta}$, $\boldsymbol{p}_{\mathrm{UE}}$, and $\boldsymbol{\xi}$, the approximate posterior
$p(\boldsymbol{x}\mid \boldsymbol{y},\boldsymbol{\delta},\boldsymbol{p}_{\mathrm{UE}},\boldsymbol{\xi})$
is computed by message passing.
Inspired by the EM framework, the proposed algorithm alternates between an E-step and an M-step until convergence.

\textbf{E-step:}
Given the uncertainty parameters $\boldsymbol{\zeta}^t$ at the $t$-th iteration, the approximate posterior
$p(\boldsymbol{x}\mid \boldsymbol{y},\boldsymbol{\zeta}^t)$
is obtained by message passing within the turbo framework.

\textbf{M-step:}
Using the approximate posterior $p(\boldsymbol{x}\mid \boldsymbol{y},\boldsymbol{\zeta}^t)$ obtained in the E-step, a surrogate function for $\ln p(\boldsymbol{y}\mid \boldsymbol{\zeta})$ is constructed and maximized with respect to $\boldsymbol{\zeta}$ via gradient ascent, yielding the next iterate $\boldsymbol{\zeta}^{t+1}$.

\subsection{E-Step}
The E-step consists of two modules, as shown in Fig.~\ref{figure_1}. Module A is an LMMSE estimator based on the observation $\boldsymbol{y}$ and the extrinsic messages from Module B. Module B performs MMSE estimation by combining the joint sparse prior with the extrinsic messages from Module A. The two modules are executed iteratively until convergence.

In Module B, message passing is used to compute the posterior distributions of $\boldsymbol{x}$ and $\boldsymbol{s}$ based on the i.i.d. channel prior and the extrinsic messages $\boldsymbol{x}_{A\rightarrow B}^{\mathrm{ext}}$ and $\boldsymbol{V}_{A\rightarrow B}^{\mathrm{ext}}$ from Module A. Specifically, the extrinsic messages are equivalently modeled by a virtual AWGN observation model:
\begin{equation}
\boldsymbol{x}_B^{\mathrm{pri}}=\boldsymbol{x}+\mathbf{z},
\end{equation}
where $\boldsymbol{x}_B^{\mathrm{pri}}=\boldsymbol{x}_{A\rightarrow B}^{\mathrm{ext}}
=\left[\left(\boldsymbol{x}_B^{1,\mathrm{pri}}\right)^T,\cdots,\left(\boldsymbol{x}_B^{K,\mathrm{pri}}\right)^T\right]^T$,
and
$\boldsymbol{V}_{A\rightarrow B}^{\mathrm{ext}}
=\operatorname{BlockDiag}\left(\boldsymbol{V}_B^{1,\mathrm{pri}},\cdots,\boldsymbol{V}_B^{K,\mathrm{pri}}\right)$
is the noise covariance, with
$\mathbf{z}^k\sim\mathcal{CN}\left(\mathbf{0},\boldsymbol{V}_B^{k,\mathrm{pri}}\right)$.

The factor graph of the corresponding joint distribution is shown in Fig.~\ref{figure_factor_graph}, and the factor functions are given by
\[
\begin{aligned}
g_q^k &\triangleq \mathcal{CN}\left(x_q^k; x_{B,q}^{k,\mathrm{pri}}, v_{B,q}^{k,\mathrm{pri}}\right), \quad \forall k,\forall q, \\
f_q^k &\triangleq p\left(x_q^k \mid s_q^k\right), \quad \forall k,\forall q, \\
\eta_q^k &\triangleq p\left(s_q^k \mid s_q\right), \quad \forall k,\forall q, \\
h_q &\triangleq p(s_q), \quad \forall q.
\end{aligned}
\]

The messages on this factor graph are derived according to the sum-product rule. First, messages are passed along the path
$x_q^k \rightarrow f_q^k \rightarrow s_q^k \rightarrow \eta_q^k \rightarrow s_q$.
Then, bidirectional message passing is performed between $s_q$ and $h_q$.
Finally, messages are propagated back through
$s_q \rightarrow \eta_q^k \rightarrow s_q^k \rightarrow f_q^k \rightarrow x_q^k$.
Since the factor graph has a tree structure, the message derivations follow directly from the sum-product rule and are omitted for brevity, as they are similar to those in~\cite{ref12}.

\subsection{M-Step}

In the M-step, directly maximizing the log-posterior is difficult because a closed-form expression for $\ln p(\boldsymbol{y},\boldsymbol{\zeta})$ is unavailable.
Instead, we construct a surrogate function for $\ln p(\boldsymbol{y}\mid \boldsymbol{\zeta})$ and maximize it with respect to $\boldsymbol{\zeta}$. At the $t$-th iteration, the surrogate function defined by the EM method is
\begin{align}
Q\left(\boldsymbol{\zeta};\boldsymbol{\zeta}^t\right)
&=\int_{\boldsymbol{x}} p\left(\boldsymbol{x}\mid \boldsymbol{y},\boldsymbol{\zeta}^t\right)
\ln \frac{p(\boldsymbol{y},\boldsymbol{x}\mid \boldsymbol{\zeta})}{p\left(\boldsymbol{x}\mid \boldsymbol{y},\boldsymbol{\zeta}^t\right)} \nonumber \\
&= -\left\|\boldsymbol{y}-\boldsymbol{\Phi}\boldsymbol{x}^{\mathrm{post}}\right\|^2
-\operatorname{tr}\left(\boldsymbol{\Phi}\boldsymbol{V}^{\mathrm{post}}\boldsymbol{\Phi}^H\right)+C,
\end{align}
where $C$ is a constant, and $\boldsymbol{x}^{\mathrm{post}}$ and $\boldsymbol{V}^{\mathrm{post}}$ are the posterior mean and covariance of $\boldsymbol{x}$ obtained in the E-step. This surrogate satisfies
\[
Q\left(\boldsymbol{\zeta};\boldsymbol{\zeta}^t\right)\leq \ln p(\boldsymbol{y},\boldsymbol{\zeta}), \quad \forall \boldsymbol{\zeta},
\]
\[
Q\left(\boldsymbol{\zeta}^t;\boldsymbol{\zeta}^t\right)=\ln p\left(\boldsymbol{y},\boldsymbol{\zeta}^t\right),
\]
and
\[
\frac{\partial Q\left(\boldsymbol{\zeta}^t;\boldsymbol{\zeta}^t\right)}{\partial \boldsymbol{\zeta}}
=
\frac{\partial \ln p\left(\boldsymbol{y},\boldsymbol{\zeta}^t\right)}{\partial \boldsymbol{\zeta}}.
\]
Therefore, the next iterate $\boldsymbol{\zeta}^{t+1}$ is obtained by maximizing
$Q\left(\boldsymbol{\zeta};\boldsymbol{\zeta}^t\right)$.
Since this function is generally non-convex, the global optimum is difficult to obtain, and gradient ascent is used for the update.

\section{Simulation Results}

In the simulations, a $300\,\text{m} \times 300\,\text{m} \times 300\,\text{m}$ 3-D area is considered. 
Three BSs are deployed at 
$[50\,\text{m}, -100\,\text{m}, 25\,\text{m}]$, 
$[50\,\text{m}, 0\,\text{m}, 25\,\text{m}]$, 
and 
$[50\,\text{m}, 100\,\text{m}, 25\,\text{m}]$, respectively. 
The user is randomly located within the considered area. The number of OFDM subcarriers is set to $N=192$, and the subcarrier spacing is $f_0=30\,\text{kHz}$. Pilot symbols are inserted every $12$ subcarriers. 
Each BS is equipped with a uniform planar array (UPA) with $N=64$ antennas ($N_x=N_y=8$). 
The standard deviation of the time offset is set to $\sigma_\tau = 1/B$ or $\sigma_\tau = 1/(2B)$, where $B=Nf_0$ denotes the total system bandwidth.

The proposed scheme is compared with the following baselines:
1) the MUSIC-LS scheme, which serves as the baseline method, 
2) the genie-aided LMMSE scheme for channel estimation performance comparison, and 
3) the ideal scheme (i.e., the proposed method without time offset) for localization performance comparison.

Fig.~\ref{figure1} shows the normalized mean square error (NMSE) of channel estimation versus the signal-to-noise ratio (SNR). 
It can be observed that the proposed scheme achieves a significantly lower NMSE than the MUSIC-LS baseline. 
This improvement demonstrates that the proposed sparse prior model effectively exploits the partially common sparsity of multi-BS channels in the location domain.

Fig.~\ref{figure2} and Fig.~\ref{figure3} illustrate the root mean square error (RMSE) of scatterer localization and user localization versus SNR, respectively. 
The proposed scheme consistently outperforms the baseline methods, particularly when the time offset is large. 
This is because the proposed framework jointly estimates the user position and time offsets by exploiting multi-BS observations, which effectively mitigates the impact of clock asynchronism.

\end{document}